\documentclass[11pt]{article}
\pdfoutput=1
\usepackage[square,comma,numbers,sort&compress]{natbib}
\usepackage{bbold}
\usepackage{amsmath}
\usepackage{amsfonts}
\usepackage{amssymb}
\usepackage{graphicx, rotating}
\usepackage{epstopdf}
\usepackage{epsfig}
\usepackage{empheq}
\usepackage{latexsym}
\usepackage{multirow}
\usepackage{stmaryrd}
\usepackage{tikz}
\usepackage{color}
\usepackage{slashed}
\usepackage[colorlinks=true,linkcolor=black,anchorcolor=black,citecolor=blue,filecolor=cyan,menucolor=red,runcolor=filecolor,urlcolor=blue,bookmarks=true,bookmarksnumbered=true]{hyperref}%
\usepackage{color}
\definecolor{rosso}{cmyk}{0,1,1,0.4}
\definecolor{rossos}{cmyk}{0,1,1,0.55}
\definecolor{rossoc}{cmyk}{0,1,1,0.2}
\definecolor{blu}{cmyk}{1,1,0,0.3}
\definecolor{blus}{cmyk}{1,1,0,0.6}
\definecolor{bluc}{cmyk}{1,1,0,0.1}
\definecolor{verde}{cmyk}{0.92,0,0.59,0.25}
\definecolor{verdec}{cmyk}{0.92,0,0.59,0.15}
\definecolor{verdes}{cmyk}{0.92,0,0.59,0.4}
\definecolor{grigio}{cmyk}{0,0,0,0.07}
\definecolor{rosa}{cmyk}{0,0.1,0.1,0.02}
\definecolor{rosino}{cmyk}{0,0.05,0.05,0.02}
\definecolor{rosas}{cmyk}{0,0.3,0.25,0.05}
\definecolor{celeste}{cmyk}{0.1,0,0,0.02}
\definecolor{giallino}{cmyk}{0,0,0.4,0.02}
\definecolor{rosso}{cmyk}{0,1,1,0.4}
\definecolor{rossos}{cmyk}{0,1,1,0.55}
\definecolor{rossoc}{cmyk}{0,1,1,0.2}
\definecolor{blu}{cmyk}{1,1,0,0.3}
\definecolor{bluc}{cmyk}{1,1,0,0.1}
\definecolor{blucc}{cmyk}{0.7,0.5,0,0}
\definecolor{viola}{cmyk}{0,1,0,0.6}
\definecolor{viola2}{cmyk}{0,1,0.2,0.6}
\definecolor{verde}{cmyk}{0.92,0,0.59,0.25}
\definecolor{verdec}{cmyk}{0.92,0,0.59,0.15}
\definecolor{verdes}{cmyk}{0.92,0,0.59,0.4}
\definecolor{verdino}{cmyk}{0.12,0,0.09,0.05}
\definecolor{giallo}{cmyk}{0,0,1,0}
\definecolor{gialloverde}{cmyk}{0.44,0,0.74,0}
\definecolor{grey}{rgb}{0.6,0.6,0.6}
\definecolor{fuchsia}{rgb}{1,0,1}

\usepackage{chngcntr}
\counterwithin{figure}{section}
\counterwithin{table}{section}
\counterwithin{equation}{section}

\def\({\left(}
\def\){\right)}
\def\be{\begin{equation}}
\def\ee{\end{equation}}
\def\bes{\begin{subequations}}
\def\ees{\end{subequations}}
\def\bea{\begin{eqnarray}}
\def\eea{\end{eqnarray}}
\def\bry{\begin{array}}
\def\ery{\end{array}}
\def\bit{\begin{itemize}}
\def\eit{\end{itemize}}
\def\ben{\begin{enumerate}}
\def\een{\end{enumerate}}
\def\l{\label}

\def\dst{\displaystyle}

\def\f{\frac}

\title{
\vspace{-1.5cm}
\normalsize{\hspace{9.5cm}DFPD-2014/TH/03, MITP/14-109}
\vspace{4cm}
\vspace{0.0 cm}
{\huge
Future tests of Higgs compositeness: direct vs indirect
}}
\vspace{0.5cm}
\author{{\Large
\text{Andrea Thamm$^{1}$\,}\footnote{athamm@uni-mainz.de}\text{\,\,,\,\,Riccardo Torre$^{2}$\,}\footnote{riccardo.torre@pd.infn.it}\text{\,\, and Andrea Wulzer$^{2}$\,}\footnote{andrea.wulzer@pd.infn.it}} \vspace{4mm}\\ 
{\small\emph{$^{1}$PRISMA Cluster of Excellence \& Mainz Institute for Theoretical Physics,}}\\ \vspace{-14pt} \\
{\small\emph{Johannes Gutenberg University, 55099 Mainz, Germany}} \\
{\small\emph{$^{2}$Dipartimento di Fisica e Astronomia, Universit\`a di Padova and}}\\ \vspace{-14pt} \\
{\small\emph{INFN, Sezione di Padova, via Marzolo 8, I-35131 Padova, Italy}} \\
}

\date{}

\begin{document}
\baselineskip=16pt

\maketitle \thispagestyle{empty}
\begin{center}
{\Large Abstract}
\end{center}
We estimate the reach of the $14\,$TeV LHC and future hadronic and leptonic colliders in the parameter space of the minimal composite Higgs model, outlining the complementarity of direct resonance searches and indirect information from the measurements of the Higgs boson couplings. The reach on electroweak charged spin--one resonances, taken here as representative direct signatures, is obtained from the current $8\,$TeV LHC limits by an extrapolation procedure which we outline and validate. The impact of electroweak precision tests, and their possible improvement at future colliders, is also quantified.

\parbox[c]{12cm}{

\medskip
\noindent

}

\newpage

\section{Introduction}
The LHC is about to restart operations at $13\,$TeV centre--of--mass energy, which will presumably be increased to $14\,$TeV within a few years. A total luminosity of $300\,$fb$^{-1}$ will be collected in the next runs, followed by a high--luminosity (HL-LHC) phase which should eventually deliver $3\,$ab$^{-1}$. While the current priority clearly lies on profiting from this experimental program, some effort should also be devoted to the design of future colliders, planning the investigation of the energy frontier on the time--scale of several decades. This may well be premature: the next LHC run could radically change the situation by discovering new particles, in which case the priority would be on characterising their properties and nature. However an assessment of future colliders' capabilities on the basis of the current theoretical understanding and experimental status might still be a useful exercise. 

Proposed future machines come in two main classes, lepton (e.g.~ILC \cite{Baer:2013cma}, CLIC \cite{Linssen:2012hp,Lebrun:2012hj}, TLEP \cite{Koratzinos:2013ncw}, also referred to as FCC-ee) and hadron (such as the FCC-hh \cite{FCC}) colliders, which will search for New Physics (NP) from complementary sides.\footnote{Here we will not consider the possibility of an electron-proton collider such as the FCC-he \cite{FCC}.} Experimental programs at lepton colliders are more suited for indirect searches, thanks to the high precision of the measurements. Hadron colliders reach higher energies and are thus more effective for direct searches of new particles. Indeed, it is not by chance that the best current indirect and direct limits on NP mostly come, respectively, from LEP and LHC data. Because of this complementarity, a comparison between the reach of lepton and hadron colliders on NP is a delicate issue, which cannot be performed in absolute terms and on completely  model--independent grounds. Some theory bias is needed, in the form of one or several NP scenarios, in order to display the reach of indirect and direct searches on the same parameter space. Here we consider the Composite Higgs (CH) scenario in its minimal realisation \cite{Contino:2003p378,Agashe:2004ib,Agashe:2005p372,Contino:2006fd,Panico:2005kd,Barbieri:2007bh,Contino:2010mr,2014arXiv1401.2457B}.

Aside from being a well--motivated theoretical possibility, CH is the ideal framework for our investigation since it predicts both indirect and direct effects which could both be sizeable enough to be detected. Telling which strategy could be more effective to test the CH idea is non--trivial and requires dedicated studies. Indirect effects, in the form of corrections to SM couplings or new BSM vertices \cite{Barbieri:2004p1607,Giudice:2007fh,Falkowski:2007kl,Low:2010hn,2011arXiv1110.5646A,Grojean:2013qca,Gupta:2014va,Montull:2013vc,Carena:2014ud,Matsedonskyi:2014wn}, unavoidably emerge due to the \sloppy\mbox{pseudo--Nambu--Goldstone} boson nature of the Higgs leading to deviations proportional to $\xi\equiv v^{2}/f^{2}$ where $f$ is the Goldstone boson Higgs decay constant and $v$ the electroweak symmetry breaking (EWSB) scale. Further corrections, but normally subdominant, come from the virtual exchange of new heavy  resonances mixing with the SM particles at tree level, giving contributions of order $m_{\text{SM}}^{2}/m_{\text{NP}}^{2}$. The latter resonances can also be produced at high enough energies, giving rise to a number of possible direct signatures. The most studied and promising ones are the production of spin--one EW--charged vectors \cite{Contino:2006fd,Agashe:2007hh,Agashe:2009bj,Agashe:2009ve,Contino:2011np,Bellazzini:2012tv,Accomando:2012us,Hernandez:2013wd,Pappadopulo:2014tg,Brooijmans:2014eja,Greco:2014aza} and of the coloured partners of the top quark (shortly referred to as top partners) \cite{DeSimone:2012ul,Matsedonskyi:2014wr,Matsedonskyi:2014wp}.

The strongest indirect constraints on CH models currently come from electroweak precision tests (EWPT), where CH models could have already shown up in the form of oblique corrections or modifications of the $Zb\bar{b}$ vertex \cite{Barbieri:2004p1607,Grojean:2013qca,Matsedonskyi:2014wn}. Even restricting to custodially symmetric cosets and to fermionic operator representations which implement the so-called $P_{LR}$ protection symmetry for $Zb\bar{b}$ \cite{Agashe:2006at}, EWPT are still the dominant indirect constraint on the CH scenario. In spite of this, and in spite of the fact that we will discuss them in detail in section \ref{EWPT}, we will not take EWPT and their possible improvements at future colliders as a central pillar of our investigation. The reason is that we judge their impact too model--dependent to be quantified in a robust way.\footnote{This is even more true for flavour constraints, which can be stronger than EWPT, but considerably more model--dependent (see, e.g., refs.~\cite{Redi:2011wr,Redi:2012gq,Barbieri:2012wia,KerenZur:2012td,DaRold:2012uc,Redi:2013vo,Grojean:2013qca,Azatov:2014wi,Antipin:2014dv,Konig:2014tt,Matsedonskyi:2014wn}).} Namely, as known in the literature and reviewed in section \ref{EWPT}, the EWPT observables are sensitive to a number of effects which can only be computed within specific and complete models and therefore are to a large extent unpredictable at the level of generality we aim to maintain here. Instead, we decided to focus on indirect effects associated to the modification of the Higgs boson couplings because they have the great virtue of being largely insensitive to many details of the specific model and thus predictable in a fairly model--independent way.\footnote{See ref.~\cite{Barbieri:2013by} for a discussion of the interplay between EWPT and Higgs coupling modifications in CH models.} This is particularly true for the trilinear Higgs coupling to EW gauge bosons which, at least for models based on the minimal coset $SO(5)/SO(4)$, is universally predicted to deviate from the SM expectation by a relative correction $k_V=\sqrt{1-\xi}$. We will thus take the sensitivity to $k_V$ of future leptonic colliders as a good model--independent measure of their reach on CH models, to be compared with direct searches at hadron colliders.

Similar considerations underly our choice of the representative direct signatures. Top partners  are very sensitive probes of CH models because their mass directly controls the generation of the Higgs potential and thus the level of fine--tuning required to achieve EWSB and a light enough Higgs boson \cite{Panico:1359049,Matsedonskyi:2012ws,Marzocca:2012tt,Panico:2012vr,Pomarol:2012vn,Pappadopulo:2013wt}. However their properties and their very existence is, to some extent, model--dependent, and we therefore do not consider top partner signatures but focus instead on EW vector resonances (see ref.~\cite{Matsedonskyi:2014wp} for a first assessment of the reach on top partners at future colliders). The existence of the latter is very robust because they are associated with the current operators of the SM group, which needs to be a global symmetry of the composite sector eventually made local by the gauging of external sources. In particular, we consider the particles associated with the SM \mbox{SU$(2)_L$} currents, which form a $(\mathbf{3},\mathbf{1})$ triplet of the unbroken strong sector group \mbox{SU$(2)_L \times$ SU$(2)_R$}. We describe this vector triplet in Model~B of ref.~\cite{Pappadopulo:2014tg}, a simplified model which depends on two parameters only: the vector triplet mass $m_\rho$ and its intrinsic coupling $g_{\rho}$ controlling the interaction with the SM fermions and the EW gauge bosons. The two parameters are related to $\xi$ by
\begin{equation}
\displaystyle
\xi=\frac{g_\rho^2}{m_\rho^2}v^2\,,
\end{equation}
from where the indirect reach on $\xi$ is immediately compared with direct searches, which set limits on the $(m_{\rho},\xi)$ or $(m_\rho,g_\rho)$ planes.

The paper is organised as follows. In section \ref{sec:extrapolation} we outline a general procedure to extrapolate resonance bounds to different energies and integrated luminosities. In section \ref{sec:results} we apply this procedure to $8\,$TeV LHC di-lepton and di-boson searches and discuss the results for the direct versus indirect reach of the $14\,$TeV LHC and future colliders. In section \ref{sec:ewpt} we provide a realistic assessment of EWPT constraints, including predictions for the improvements at ILC and TLEP, by taking the aforementioned model--dependent effects into account. Finally in section \ref{sec:conclusion} we report our conclusions. In the appendix we present a simple check of the extrapolation procedure outlined in section \ref{sec:extrapolation} and discuss its range of validity.  Some of these results were presented by one of us in a preliminary version in ref.~\cite{Thamm:2014fba}.

\section{Limit extrapolation}\l{sec:extrapolation}

Based on the $8$\,TeV LHC data, the ATLAS and CMS collaborations have performed a number of vector resonance searches in different final states, setting limits on the production cross section times branching ratio as a function of the resonance mass $m_\rho$. We thus have a set of \mbox{$[\sigma\hspace{-3pt}\times\hspace{-3pt} \text{BR}](s_0,L_0;m_\rho)$} curves in the different search channels, obtained at a centre-of--mass energy of $\sqrt{s_0}=8\,$TeV and with an integrated luminosity \mbox{$L_0\simeq20\,$fb$^{-1}$}. We now describe a strategy to extrapolate these limits to a different proton--proton collider of energy $\sqrt{s}$ and luminosity $L$, producing \mbox{$[\sigma\hspace{-3pt}\times\hspace{-3pt} \text{BR}](s,L;m_\rho)$} curves. This procedure delivers exclusion limits, obtained in the absence of any signal, which can however also be regarded as estimates of the future colliders' sensitivity at the level of approximation we are working here.

The basic idea underlying our extrapolation is that the limit is essentially driven by the number $B(s,L,m_\rho)$ of background events which are present, for a given collider configuration, in a  small window of partonic invariant mass squared $\hat{s}$ (of fixed relative width \mbox{$\Delta{\hat{s}}/m_\rho^2\ll1$}) centred around the resonance mass. Our assumption means that the upper limit on the number of signal events at each mass point, from which the excluded \mbox{$[\sigma\hspace{-3pt}\times\hspace{-3pt} \text{BR}]$} is obtained at a given luminosity, is exclusively a function of the estimated number of background events from which the excluded signal is statistically extracted. Clearly this only holds up to the signal acceptance and efficiency which we consider to be fairly independent of the resonance mass and collider energy. Now we can define an ``equivalent mass'' $m_\rho$ for each resonance mass $m_\rho^0$ on the $8\,$TeV exclusion plot, as the mass with the same number of associated background events at the new collider energy and luminosity. Namely, we obtain $m_\rho$ by inverting the equation
\begin{equation}\label{eq5}
B(s,L,m_\rho)=B(s_0,L_0,m_\rho^0)\,.
\end{equation}
For each given $m_\rho^0$, the associated equivalent mass $m_\rho$ is by definition the one characterised by having the same number of background events in the search region. According to the previous discussion, it therefore gives rise to the same limit on the number of signal events. The excluded cross--section at the equivalent mass is thus obtained from the $8\,$TeV limit by rescaling the integrated luminosity \footnote{Notice that the acceptance times efficiency factor, which enters in the relation between the number of excluded signal events and the excluded cross--section, cancels because we assumed it to be constant.}
\begin{equation}
\label{eql}
\displaystyle
[\sigma\hspace{-3pt}\times\hspace{-3pt} \text{BR}](s,L;m_\rho)=\frac{L_{0}}{L}\cdot[\sigma\hspace{-3pt}\times\hspace{-3pt} \text{BR}](s_0,L_0;m_\rho^0)\,.
\end{equation}
Extracting the equivalent $m_\rho$ defined by eq.~(\ref{eq5}) and applying the equation above for each value of $m_\rho^0$, we can extrapolate the $8\,$TeV limits to any collider energy and integrated luminosity.

Before describing the procedure in detail, it is worth warning the reader that our assumptions are rather strong and not necessarily very accurate. In particular the fact that the limit is driven by the background around the peak is only strictly true if the search is performed as a counting experiment of the events falling into a window around the resonance mass. However, this is not what is done at the LHC at $8\,$TeV and will be done at future colliders. Shape analyses are performed to improve the reach and, a priori, the cross--section limits depend on background and signal kinematical distributions in a non--trivial way. However, we make the reasonable assumption that the final result is actually not far from the one obtainable with a cut--and--count strategy, which we expect to be the case within a factor of a few on the  \mbox{$[\sigma\hspace{-3pt}\times\hspace{-3pt} \text{BR}]$} reach. In the simple case of di-lepton searches, such as those of refs.~\cite{ATLAScollaboration:2014eb, CMS-PAS-EXO-12-061}, we verified that this is actually true within a factor of two for a window of relative size \mbox{$\Delta{\hat{s}}/m_\rho^2=10\%$} and for narrow resonances, but larger corrections might arise in other cases. The limits presented here should thus be regarded as ${\mathcal{O}}(1)$ estimates. However they are accurate enough for the current stage of future colliders studies.

In order to determine the equivalent mass defined in eq.~(\ref{eq5}) we proceed as follows. The number of background events is given by
\begin{equation}\label{eq1}
B(s,L,m_\rho)\propto L\cdot\sum_{\left\{i,j\right\}}\int{d\hat{s}}\frac{1}{\hat{s}}\frac{d{\mathcal{L}}_{ij}}{d\hat{s}}(\sqrt{\hat{s}};\sqrt{s})\left[\hat{s}\hat{\sigma}_{ij}\left(\hat{s}\right)\right]\,,
\end{equation}
where the integral is performed in the window ${\hat{s}}\in[m_\rho^2-\Delta{\hat{s}}/2,m_\rho^2+\Delta{\hat{s}}/2]$ according to our assumption. In the equation, $d{\mathcal{L}}_{ij}/d\hat{s}$ denotes the parton luminosity of each partonic channel ${i,j}$ which we sum over, defined as
\begin{equation}\label{eq2}
\frac{d{\mathcal{L}}_{ij}}{d\hat{s}}(\sqrt{\hat{s}};\sqrt{s})=\frac{1}{s}\int_{\hat{s}/s}^{1}\frac{dy}{y}f_{i}\left(y ;\,\hat{s}\right)f_{j}\left(\frac{\hat{s}}{y\,s};\,\hat{s}\right)\,,
\end{equation}
in terms of the parton distribution functions $f_{i}(x\,,Q^2)$ evaluated at the factorisation scale $Q^2=\hat{s}$. The parton luminosity depends both on the collider centre--of--mass energy $\sqrt{s}$ and on the partonic one $\sqrt{\hat{s}}$. The cross--section of the partonic reactions contributing to the background are denoted by $\hat{\sigma}_{ij}$ in eq.~(\ref{eq1}). Since they describe SM processes at energies much above the SM masses, they show a scale--invariant behaviour at tree--level, i.e.
\begin{equation}
\displaystyle
\left[\hat{s}\hat{\sigma}_{ij}\left(\hat{s}\right)\right]\simeq c_{ij}\,,
\end{equation}
where $c_{ij}$ are process--dependent constants. In our assumption, the background is restricted to a narrow window \mbox{$\Delta{\hat{s}}\ll m_\rho^2$} so that the parton luminosities are nearly constant in the integration region and our background prediction becomes
\begin{equation}\label{eq3}
B(s,L,m_\rho)\propto \frac{\Delta{\hat{s}}}{m_\rho^2}\cdot
L\cdot\sum_{\left\{i,j\right\}}c_{ij}\frac{d{\mathcal{L}}_{ij}}{d\hat{s}}(m_\rho;\sqrt{s})\,.
\end{equation}
By equating the backgrounds as prescribed by eq.~(\ref{eq5}) the relative width \mbox{$\Delta{\hat{s}}/m_\rho^2$} and the other pre--factors cancel and we obtain
\begin{equation}
\displaystyle
\label{eq6}
\sum_{\left\{i,j\right\}}c_{ij}\frac{d\mathcal{L}_{ij}}{d\hat{s}}(m_\rho;\sqrt{s})=\frac{L_{0}}{L}\sum_{\left\{i,j\right\}}c_{ij}\frac{d\mathcal{L}_{ij}}{d\hat{s}}({m_\rho^0};\sqrt{s_0})\,.
\end{equation}

\begin{figure*}[t!]
\begin{center}
\vspace{-2cm}\includegraphics[scale=0.55]{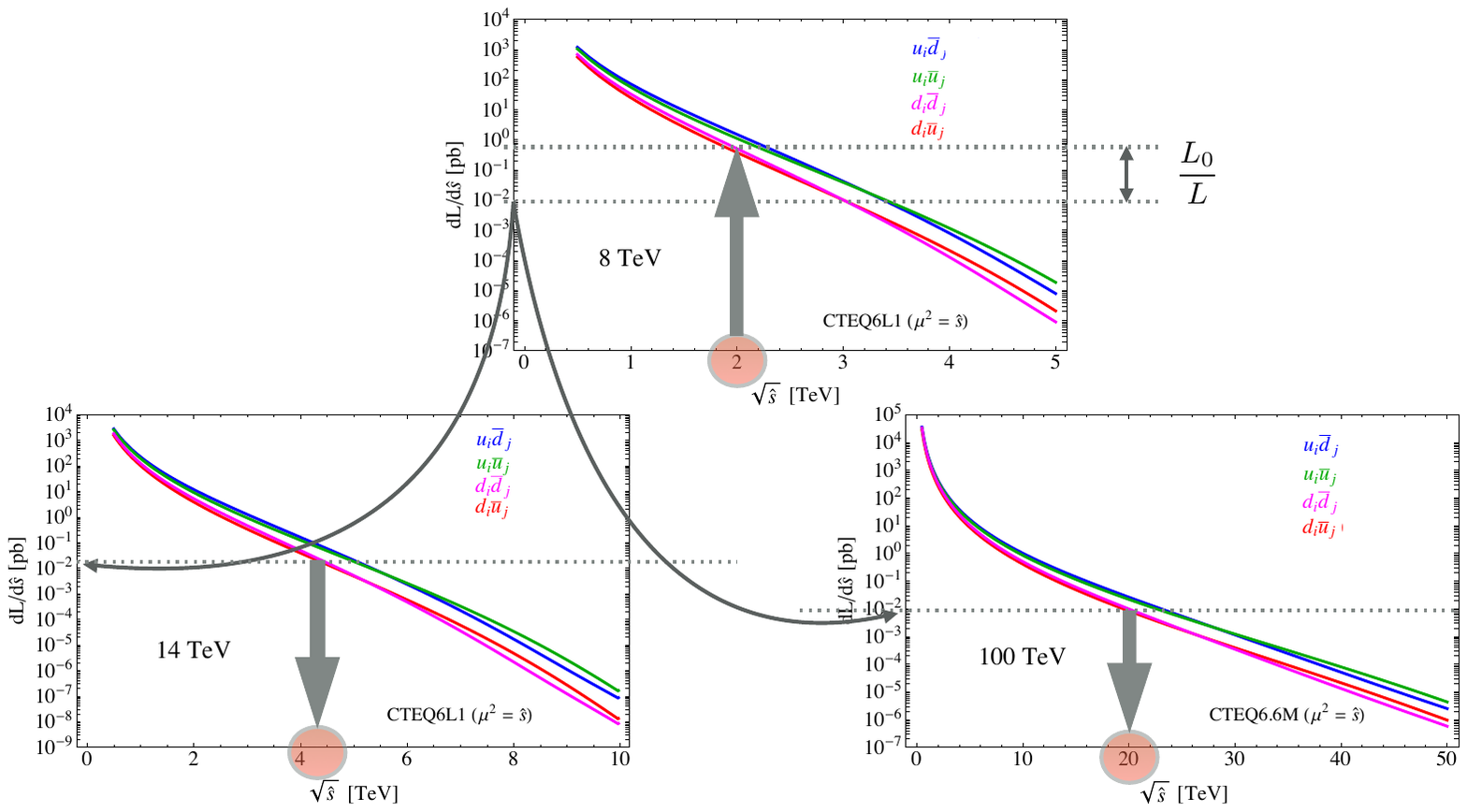}\vspace{-2cm}
\caption{\small Sketch of the procedure used to extrapolate bounds to different collider configurations.}\label{fig:extrapolatingprocedure}\vspace{-0.5cm}
\end{center}
\end{figure*}

The extrapolation procedure is depicted in figure \ref{fig:extrapolatingprocedure}. 
For each search channel we first have to identify the relevant background processes with the associated parton luminosities. The simplest case is a background dominated by a single partonic initial state where the sum drops in eq.~(\ref{eq6}), but also the case of a mixed background composition is easy to deal with. In the first case the relevant parton luminosity is the one of the dominant background process, in the second one what matters is a linear combination of the parton luminosities in the different channels with coefficients $c_{ij}$ (possibly normalised to unity). At each mass $m_\rho^0$ we first identify the relevant parton luminosity function at the $8\,$TeV LHC, read its value at $\sqrt{\hat{s}}=m_\rho^0$ and rescale it by the luminosity ratio $L_{0}/L$.  We then take the parton luminosity at the new collider energy $\sqrt{s}$, e.g.~$14$ or $100\,$TeV as depicted in the figure, and evaluate the mass where it equals the rescaled $8\,$TeV value previously determined. According to eq.~(\ref{eq6}) this delivers the equivalent mass $m_\rho$ associated with $m_\rho^0$, where the cross--section limit is provided by eq.~(\ref{eql}). 

The extrapolated limits could be obtained by applying the described procedure for each value of $m_\rho^0$ covered in the $8\,$TeV exclusion plot. However, we alter the procedure slightly due to the following subtlety. 
The $8\,$TeV exclusion plots extend over a finite mass range with the lowest mass point $(m_\rho^0)_{\textrm{min}}$ determined by the sensitivity of the specific analysis. The equivalent mass associated to this minimal $(m_\rho^0)_{\textrm{min}}$ is the lowest one which we would obtain by the extrapolation with a fixed integrated luminosity and would therefore set the lowest mass point in the extrapolated curve. The starting point of the extrapolated plot would therefore become arbitrary depending on the considered integrated luminosity. Furthermore, the lowest equivalent $(m_\rho)_{\textrm{min}}$ mass obtained from $(m_\rho^0)_{\textrm{min}}$ grows with the luminosity of the new collider, so that the exclusion limit starts at a higher mass for higher luminosity. This would lead to the paradoxical situation where some mass points could be excluded only with a smaller amount of collected data. Moreover, mass--points which were too low to be relevant for the $8\,$TeV search might end up in a relevant signal region after extrapolation.
We solve this problem by smoothly raising the integrated luminosity of the new collider up to the desired total $L$, drawing the extrapolated limits by taking the strongest at each mass. Above the value of $(m_\rho)_{\textrm{min}}$ the strongest bound comes from the highest integrated luminosity $L$, while below that it comes from a lower luminosity. The low--mass limit is thus conservative and not optimal, as it would be obtainable with a smaller set of data. This is verified explicitly in the appendix, where a validation of the extrapolation procedure is presented in the case of di-lepton resonance searches.

\section{Results}\l{sec:results}

\begin{figure*}[t!]
\begin{center}
\includegraphics[scale=0.235]{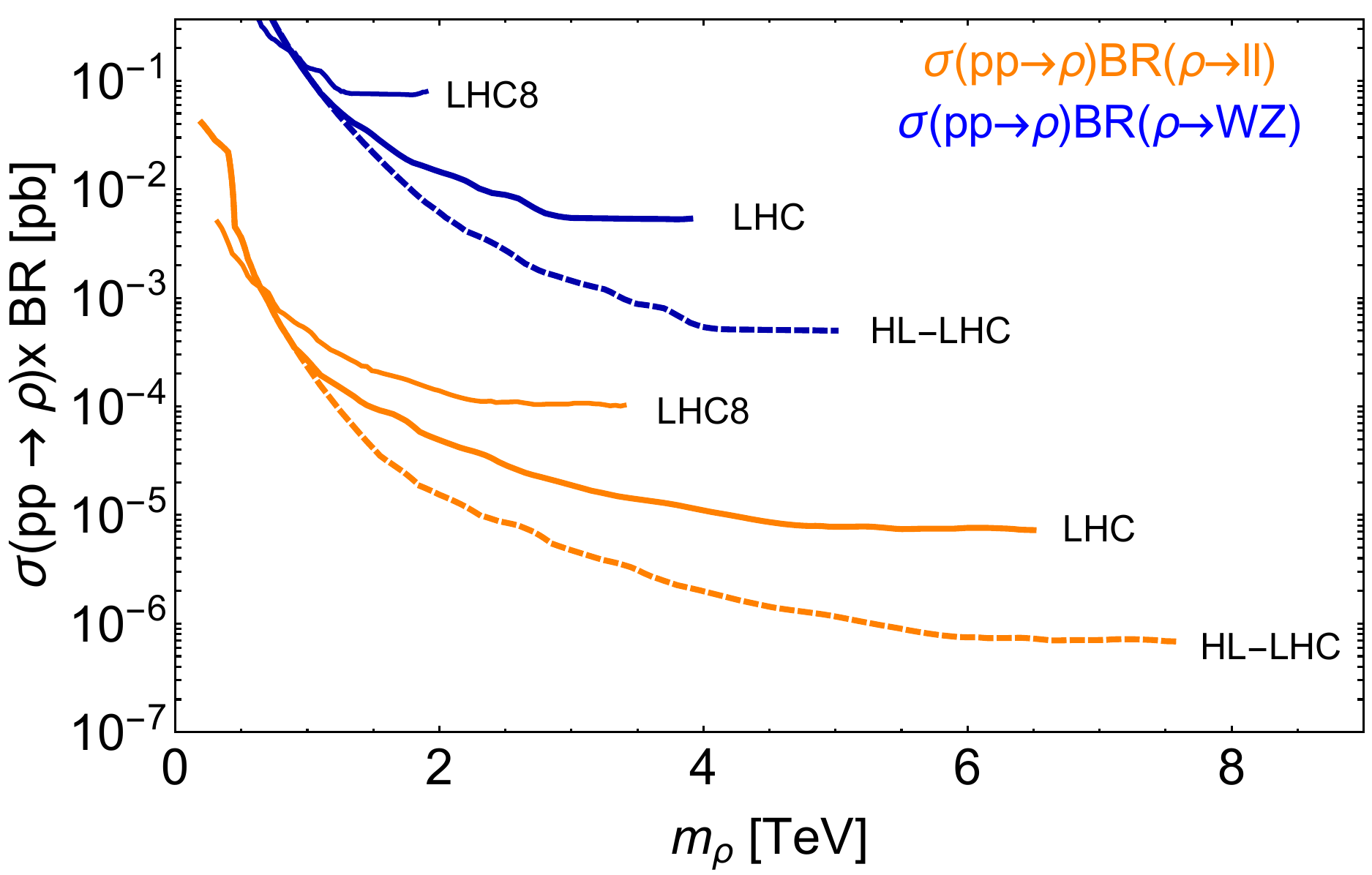}
\hspace{0.5cm}\includegraphics[scale=0.235]{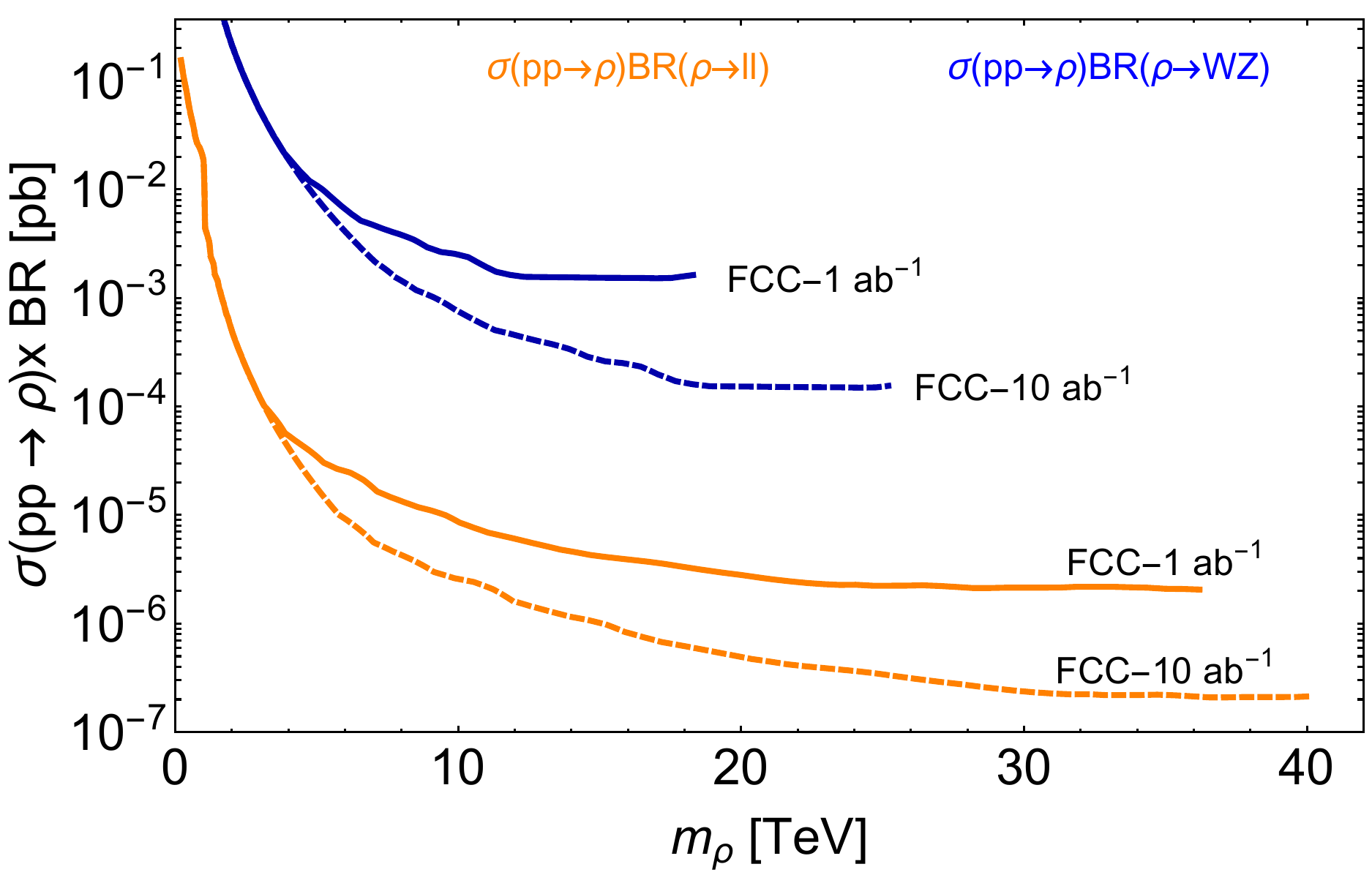}\vspace{-0.2cm}
\caption{\small Bounds on $\sigma\times \text{BR}$ from LHC at $8\,$TeV (LHC8) with $20\,$fb$^{-1}$ (solid) and corresponding extrapolations to LHC at $14\,$TeV with $300\,$fb$^{-1}$ (solid) (LHC) and $3\,$ab$^{-1}$ (dashed) (HL-LHC) and to FCC at $100\,$TeV with $1\,$ab$^{-1}$ (solid) and $10\,$ab$^{-1}$ (dashed). The two analyses of refs.~\cite{CMS-PAS-EXO-12-061} (CMS di-leptons, orange) and \cite{Khachatryan:2014xja} (CMS fully leptonic di-bosons, blue) are considered.}\label{fig:extrapolatedbounds}\vspace{-0.4cm}
\end{center}
\end{figure*}

Figure~\ref{fig:extrapolatedbounds} shows the current $8\,$TeV LHC limits with $20\,$fb$^{-1}$ ($95\%\,$CL expected exclusions) on $\sigma\times \text{BR}$, used as inputs, and the extrapolated bounds at the $14\,$TeV LHC and the $100\,$TeV FCC with integrated luminosities of $300\,$fb$^{-1}$ and $3\,$ab$^{-1}$ and $1\,$ab$^{-1}$ and $10\,$ab$^{-1}$, respectively. For definiteness, we restrict our attention to the CMS search for opposite sign di-leptons in ref.~\cite{CMS-PAS-EXO-12-061} and for fully leptonic $WZ$ in ref.~\cite{Khachatryan:2014xja}.\footnote{In the experimental analyses that we consider, the leptonic branching ratios of the bosons are defined as the average of the BRs into electrons and muons. Therefore one has, for instance, $\text{BR}(V\to ll)=1/2\(\text{BR}(V\to e^{+}e^{-})+\text{BR}(V\to \mu^{+}\mu^{-})\)$.} We verified that the corresponding ATLAS results in refs.~\cite{ATLAScollaboration:2014eb} and \cite{ATLAScollaboration:2014uc} yield similar limits. Searches for other final states could be considered as well but would not change the picture qualitatively.\footnote{See refs.~\cite{Pappadopulo:2014tg} and \cite{BridgeToAppear} for a complete list of $8\,$TeV heavy vector searches.} Notice that the di-lepton and $WZ$ channels are respectively sensitive to the electrically neutral and charged components of the triplet. The limits in the neutral and charged channels are easily compared since the properties of the two states (namely masses, production rates and Branching Ratios) are tightly related in a model--independent way \cite{Pappadopulo:2014tg}. Furthermore notice, that considering a leptonic and bosonic channel ensures an appropriate coverage of the model parameter space: the di-lepton channel dominates for small coupling $g_\rho$ while di-bosons become relevant at large $g_\rho$, where the leptonic BR deteriorates.

The limits in figure \ref{fig:extrapolatedbounds} show a number of expected features. First, they approach constants at large masses, corresponding to the cross--section limit set by zero background events. These horizontal asymptotes could safely be extended to infinite masses provided that the background decreases monotonically. However the limits above the high--mass endpoint of the curves obtained by the extrapolation are not relevant since our signal cross--section is never large enough at such high masses. We also notice that a luminosity upgrade by a factor of ten (from $300\,$fb$^{-1}$ to $3\,$ab$^{-1}$ at the LHC or from $1\,$ab$^{-1}$ to $10\,$ab$^{-1}$ at the FCC) correctly improves the cross--section reach by one order of magnitude in the high mass region while the relative improvement reduces to around three when going to lower masses and entering the region where background becomes considerable. This feature disappears at even lower masses, where the two luminosity curves start to coincide. This is due to the fact that our extrapolation procedure at low masses is unreliable as we described above and will detail in the appendix. Finally, we observe that the $14\,$TeV  LHC limits at relatively low masses are weaker than the corresponding $8\,$TeV ones and a similar situation is encountered in the comparison between the FCC and the LHC. This is due to the much larger background expected at a collider of higher energy at low masses. However the growth of the signal cross--section will overcompensate this effect and the higher energy collider eventually leads to stronger limits in the entire relevant mass range as we show below.

\begin{figure*}[t!]
\begin{center}
\includegraphics[scale=0.23]{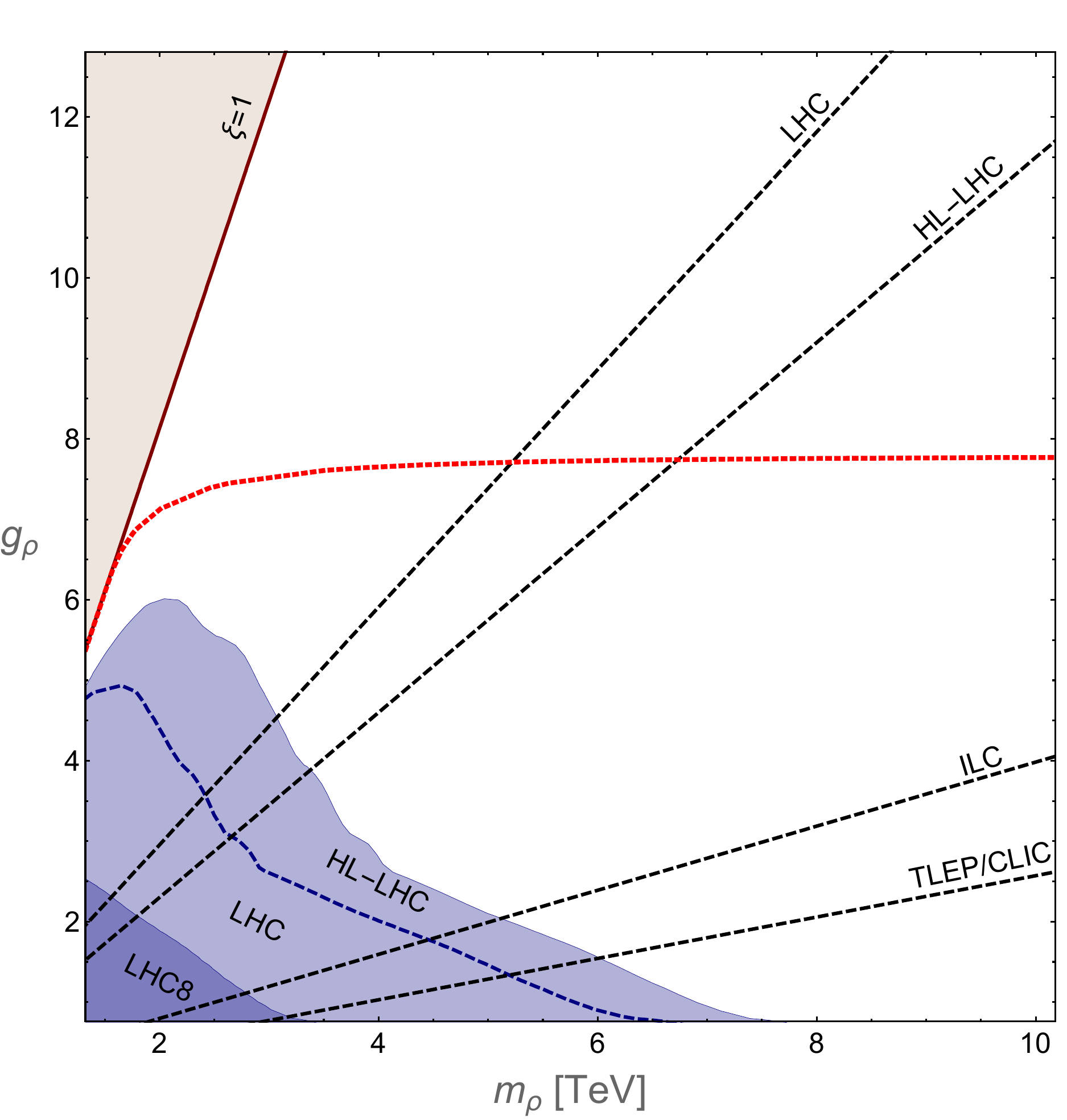}\hspace{4mm}
\hspace{0.5cm}\includegraphics[scale=0.23]{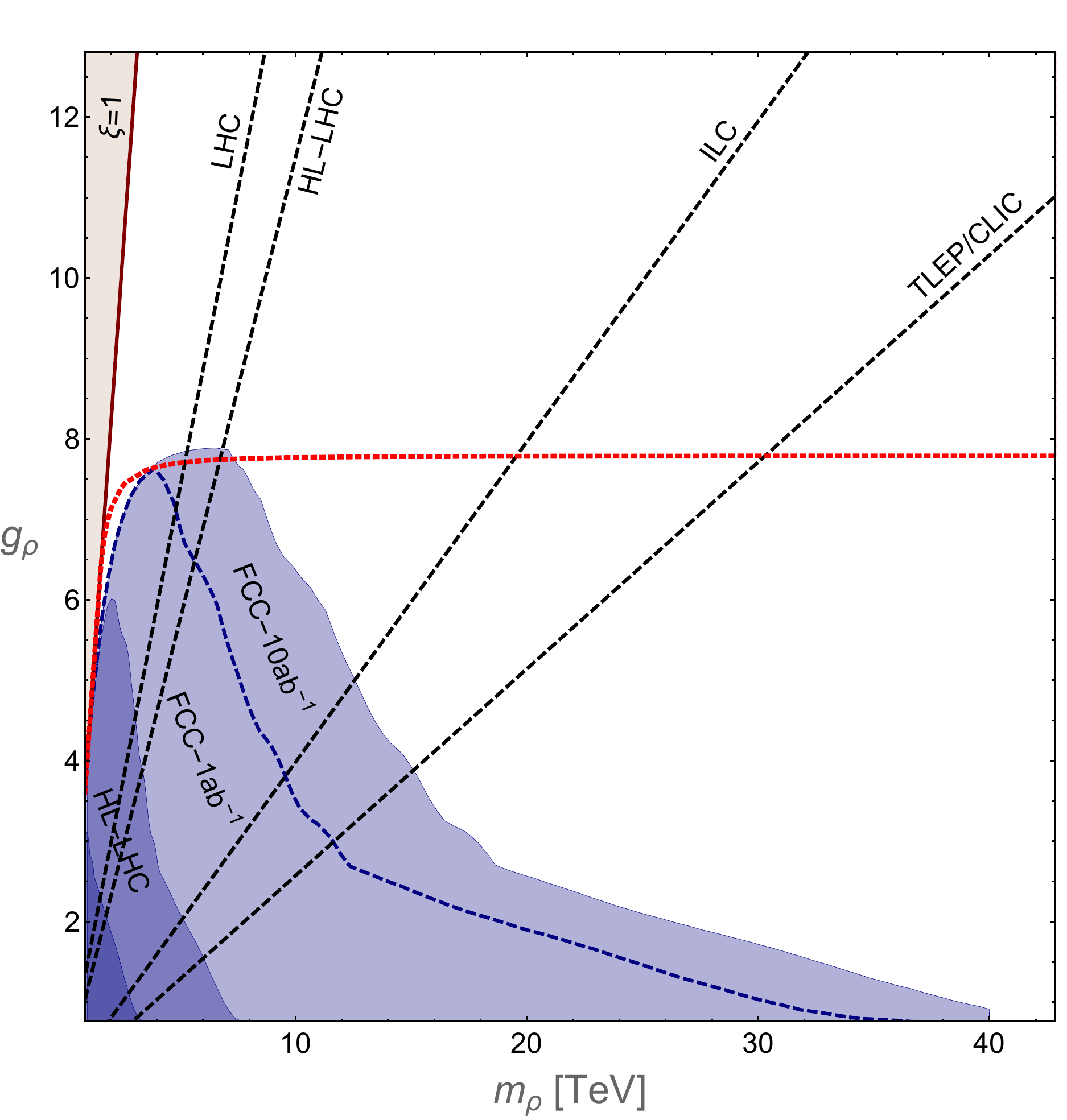}
\caption{\small Comparison of direct and indirect searches in the $(m_{\rho},g_{\rho})$ plane. Left panel: region up to $m_{\rho}=10\,$TeV showing the relevance of LHC direct searches at $8\,$TeV with $20\,$fb$^{-1}$ (LHC8), $14\,$TeV with $300\,$fb$^{-1}$ (LHC) and $3\,$ab$^{-1}$ (HL-LHC); right plot: region up to $m_{\rho}=40\,$TeV showing the comparison between the LHC and FCC reach with $1$ and $10\;$ab$^{-1}$. Indirect measurements at the LHC, HL-LHC, ILC at $500\,$GeV with $500\,$fb$^{-1}$ and TLEP at $350\,$GeV with $2.6\,$ab$^{-1}$ are shown.}\label{fig:mrho_grho_plane}\vspace{-0.5cm}
\end{center}
\end{figure*}

\begin{figure*}[t!]
\begin{center}
\includegraphics[scale=0.23]{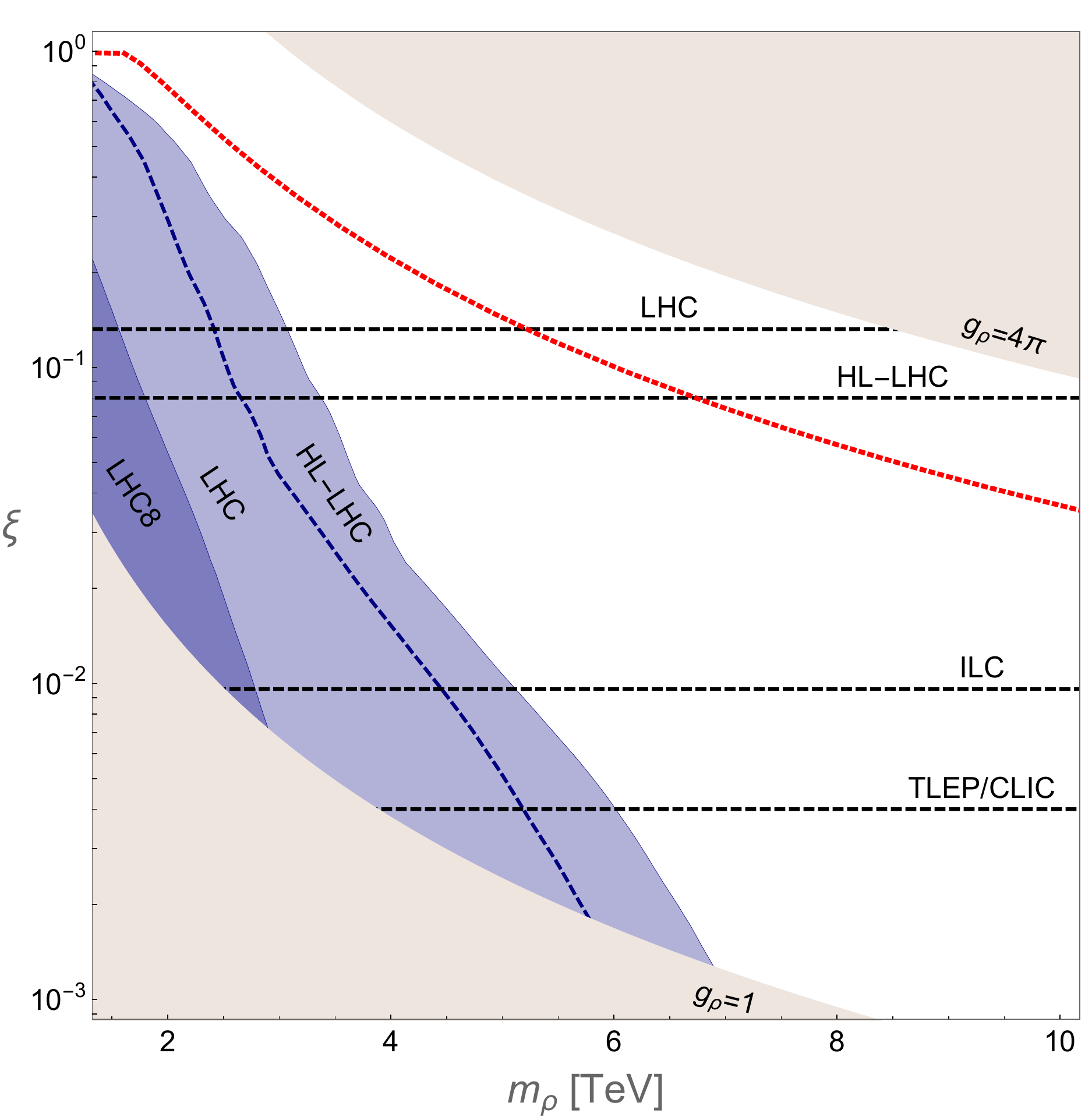}\hspace{2mm}
\hspace{0.5cm}\includegraphics[scale=0.23]{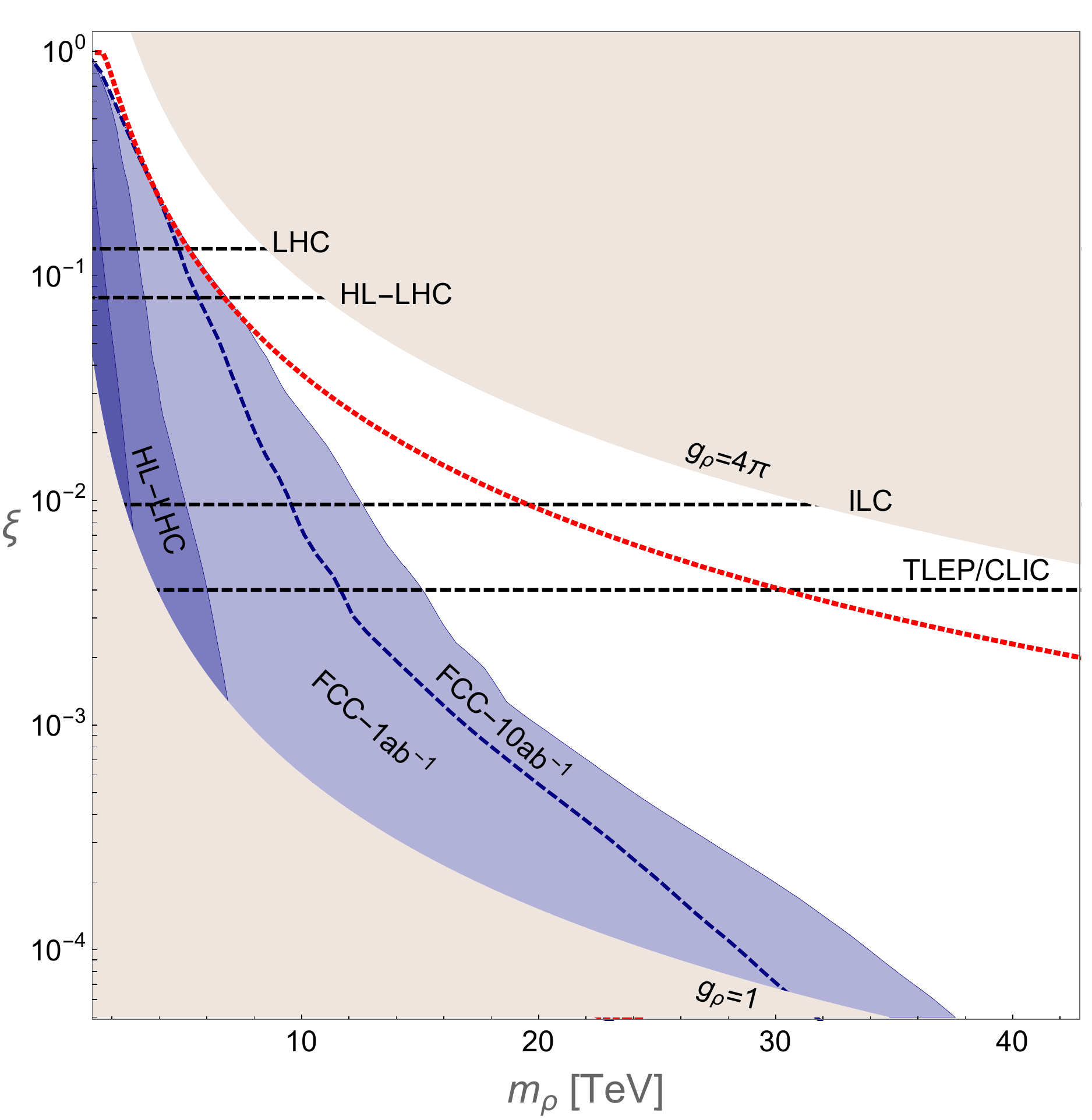}
\caption{\small Comparison of direct and indirect searches in the $(m_{\rho},\xi)$ plane. Left panel: region up to $m_{\rho}=10\,$TeV showing the relevance of LHC direct searches at $8\,$TeV with $20\,$fb$^{-1}$ (LHC8), $14\,$TeV with $300\,$fb$^{-1}$ (LHC) and $3\,$ab$^{-1}$ (HL-LHC); right plot: region up to $m_{\rho}=40\,$TeV showing the comparison between the LHC and FCC reach with $1$ and $10\;$ab$^{-1}$. Indirect measurements at the LHC, HL-LHC, ILC at $500\,$GeV with $500\,$fb$^{-1}$ and TLEP at $350\,$GeV with $2.6\,$ab$^{-1}$ are shown.}\vspace{-0.5cm}\label{fig:mrho_xi_plane}
\end{center}
\end{figure*}

The bounds on $\sigma\times \text{BR}$ shown in figure \ref{fig:extrapolatedbounds} can be translated into $95\%\,$CL allowed and excluded regions in the parameter space of our simplified model. The results are shown in figures~\ref{fig:mrho_grho_plane} and \ref{fig:mrho_xi_plane} in the $(m_{\rho},g_{\rho})$ and  $(m_{\rho},\xi)$ planes. The left panels of the two figures depict the region relevant for the LHC, while the right panels show the full reach of the FCC at $100$\;TeV. The viable region of the CH parameter space constrains $g_\rho$ to be stronger than the SM couplings but still within the perturbative regime, $1 \leq g_\rho \leq 4\pi$, and $\xi \le 1$. The regions which violate these conditions are theoretically excluded and coloured in grey in the plots. The color convention which we adopt in both figures is as follows. Violet shaded regions are excluded by direct searches at different collider configurations, starting from the LHC at $8$\;TeV and $20\;$fb$^{-1}$ (darkest), the high luminosity LHC at $14\;$TeV with $3\;$ab$^{-1}$ (medium dark) and the FCC with $10\;$ab$^{-1}$ (lightest). The violet dashed lines represent the $14\;$TeV LHC with $300\,$fb$^{-1}$ in the left plots and the FCC with $1\,$ab$^{-1}$ in the right ones. 

The shape of the limits in figure \ref{fig:mrho_grho_plane} is easily understood by simple physical considerations \cite{Pappadopulo:2014tg}. Due to partial compositeness the coupling to fermions scales as $1/g_{\rho}$ and thus the Drell-Yan production cross section, which is by far the dominant channel, decreases as $1/g_{\rho}^{2}$ in the large--coupling limit. In a somewhat counterintuitive way, the resonance becomes effectively weakly--coupled at large $g_{\rho}$ and this is why the mass--reach deteriorates. The presence of a kink in the limits originates from the superposition of the di-lepton and di-boson searches we considered which, as already mentioned, is more sensitive to weak and strong $g_\rho$, respectively. This is due to the fact that, while the coupling to fermions decreases, the one to (longitudinal) gauge bosons increases like $g_{\rho}$ and the di-boson BR rapidly becomes dominant. 

The global message which emerges from these pictures is rather simple and expected. An increase of the collider energy improves the mass reach dramatically, and in particular only the $100\,$TeV FCC can access the multi--TeV region. An increase in luminosity, instead, has a marginal effect on the mass reach but considerably extends the sensitivity in the large $g_\rho$ (i.e., small rate) direction. In particular we see that the impact of the high luminosity extension of the LHC is considerable given that largish values of the $g_\rho$ coupling are perfectly plausible in the CH scenario (see the Conclusions for a more detailed discussion).

Let us now turn to the indirect constraints from the measurement of the Higgs coupling to vector bosons. The $1\,\sigma$ ($68\%\,$CL) error on $\xi$ (i.e., twice the one on $k_V\simeq 1-\xi/2$) obtainable for different collider options, as extracted from currently available literature, are summarised in table~\ref{tab:reachoncomp}. Twice those values, which in the assumption of gaussian statistics corresponds to the $95\%\,$CL limits on $\xi$, are reported in figures~\ref{fig:mrho_grho_plane} and \ref{fig:mrho_xi_plane} as black dashed curves, with the excluded region sitting above the lines. In the $(m_{\rho},\xi)$ plane, the limits simply corresponds to horizontal lines and translate into straight lines with varying inclination in the $(m_{\rho},g_\rho)$ plane. In particular, we show the LHC reach with $300\,$fb$^{-1}$ and $3\,$ab$^{-1}$, obtained from single Higgs production, corresponding to $\xi > 0.13$ and $\xi>0.08$ respectively, and the expected reach of the ILC and TLEP at $\sqrt{s} = 500\,$GeV and $\sqrt{s} = 350$\,GeV corresponding to $\xi>0.01$ and $\xi>0.004$. Note that CLIC with 2 ab$^{-1}$ is expected to have a sensitivity comparable to TLEP.

We can now appreciate the complementarity of direct and indirect searches in exploring the parameter space of the CH scenario: direct searches are more effective for small $g_\rho$ while indirect measurements win in the large coupling region. At the LHC with $300\,$fb$^{-1}$ direct searches will completely cover the region accessible by indirect measurements at the same collider for $g_\rho \lesssim 3.5$ and it is only for $g_\rho> {g_\rho}^{\textrm{max}}=3.5$ that the latter will explore novel territory. Since direct and indirect constraints benefit similarly from the luminosity improvement, the $g_\rho^{\textrm{max}}$ threshold remains unchanged at the HL--LHC. As far as future machines are concerned, $g_\rho^{\textrm{max}}\simeq 4$ in the comparison between the $10\,\text{ab}^{-1}$ FCC and TLEP and  $g_\rho^{\textrm{max}}\simeq 5$ for FCC versus ILC. On the other hand direct searches become ineffective at large coupling, not only because of the reduction of the production cross--section as explained above but also for the following reason. An effect, which is not taken into account in our analysis, is that the resonances become broad for large $g_\rho$ because their coupling to longitudinal vector bosons and Higgs grows, increasing the intrinsic width as $g_\rho^2$. Broad resonances are harder to see and since a narrow resonance has been assumed in our analysis we expect the actual limits to be even weaker than ours in the large coupling regime. One can get an idea of where finite width effects should start to become relevant and our estimates might fail by looking at the fine red dotted curves which are shown in all plots. Above this bound the total resonance width exceeds $20\%$ of the mass and our bounds are not reliable anymore (see ref.~\cite{Pappadopulo:2014tg} for a more quantitative assessment of the width effects).

\begin{table}[t!]
\begin{center}
{\small
\begin{tabular}{ll|llc}
Collider 	& \phantom{+} Energy 		& \ Luminosity		 		& $\xi \,\,[1\sigma]$ 						& References  \\[0.2cm]\hline\hline
LHC  	& \phantom{+} $14\,$TeV 		& \ $300\,\text{fb}^{-1}$ 		& $6.6 - 11.4 \times 10^{-2}$  			& \cite{CMS-NOTE-2012/006,ATL-PHYS-PUB-2013-014,Dawson:2013bba} \\[0.2cm] \hline
LHC  	& \phantom{+} $14\,$TeV 		& \ $3\,\text{ab}^{-1}$ 		& $4 - 10 \times 10^{-2}$  			& \cite{CMS-NOTE-2012/006,ATL-PHYS-PUB-2013-014,Dawson:2013bba} \\[0.2cm] \hline
ILC  		& \phantom{+} $250\,$GeV  	& \ $250\,\text{fb}^{-1}$	        & \multirow{2}{*}{4.8-7.8$\, \times 10^{-3}$} 	& \multirow{2}{*}{\cite{Baer:2013cma,Dawson:2013bba}} \\[0.05cm]
		& + $500\,$GeV  			& \ $500\,\text{fb}^{-1}$ 		& \\[0.2cm]\hline
CLIC 	& \phantom{+} $350\,$GeV \ 	& \ $500\,\text{fb}^{-1}$		& \multirow{3}{*}{2.2 $\, \times 10^{-3}$} 	& \multirow{3}{*}{\cite{Abramowicz:2013tzc,Dawson:2013bba}} \\[0.05cm]
		& + $1.4\,$TeV 				& \ $1.5\,\text{ab}^{-1}$ 		& \\[0.05cm]
		& + $3.0\,$TeV  			& \ $2\,\text{ab}^{-1}$ 		& \\[0.2cm]\hline
TLEP 	& \phantom{+} $240\,$GeV \ 	& \ $10\,\text{ab}^{-1}$ 		& \multirow{2}{*}{2$\, \times 10^{-3}$} 		& \multirow{2}{*}{\cite{Dawson:2013bba}}\\[0.05cm] 
		& + $350\,$GeV &\ $2.6\,\text{ab}^{-1}$ & \\[0.2cm]
\end{tabular}
}
\\[0.1cm]
\caption{\small 
Summary of the reach on $\xi$ (see the text for the definition) for various collider options.
}\vspace{-0.4cm}
\label{tab:reachoncomp}
\end{center}
\end{table}

\section{EWPT reassessment}
\begin{figure*}[t!]
\begin{center}
\includegraphics[scale=0.35]{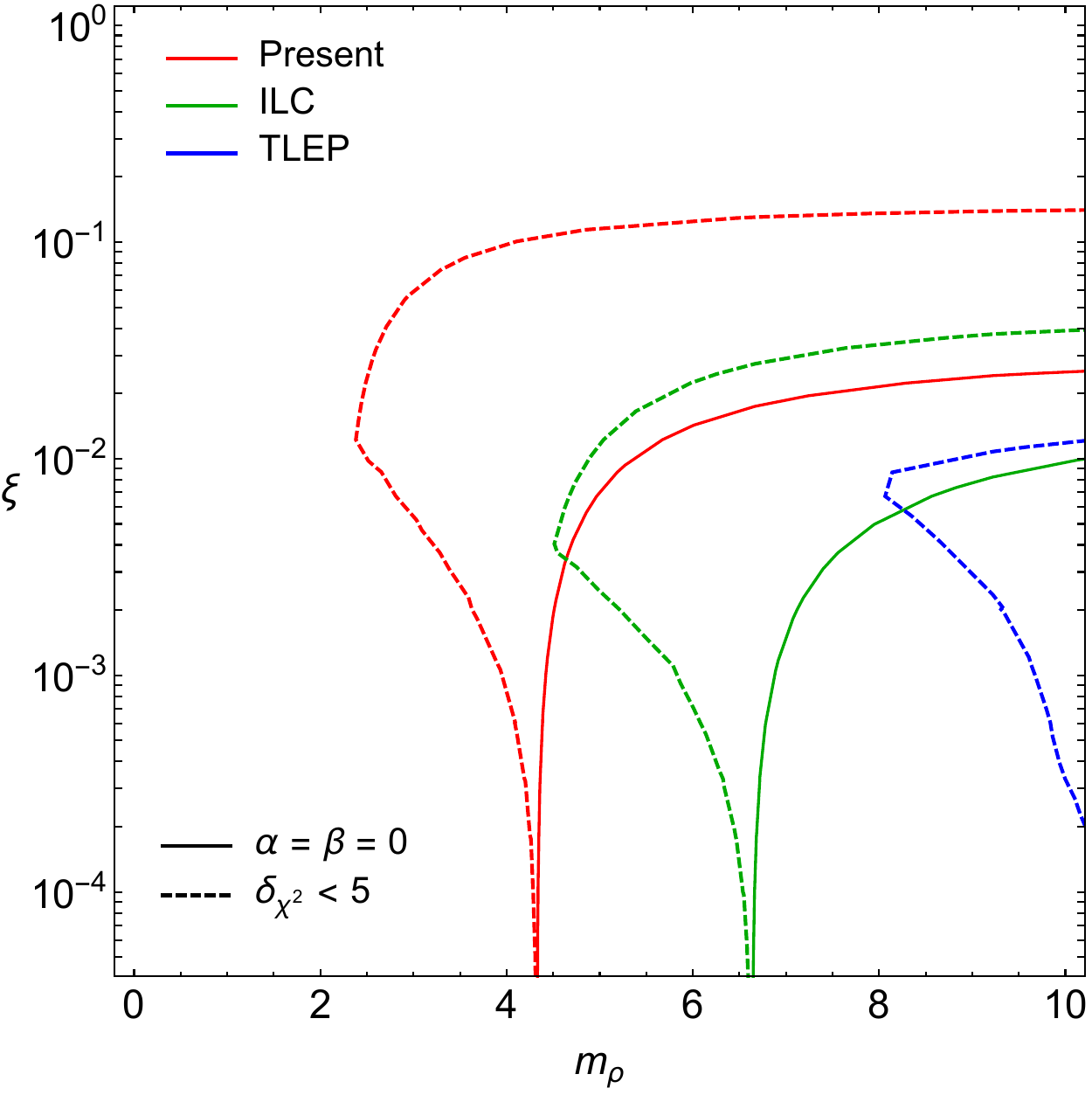}\hspace{2mm}
\hspace{0.5cm}\includegraphics[scale=0.35]{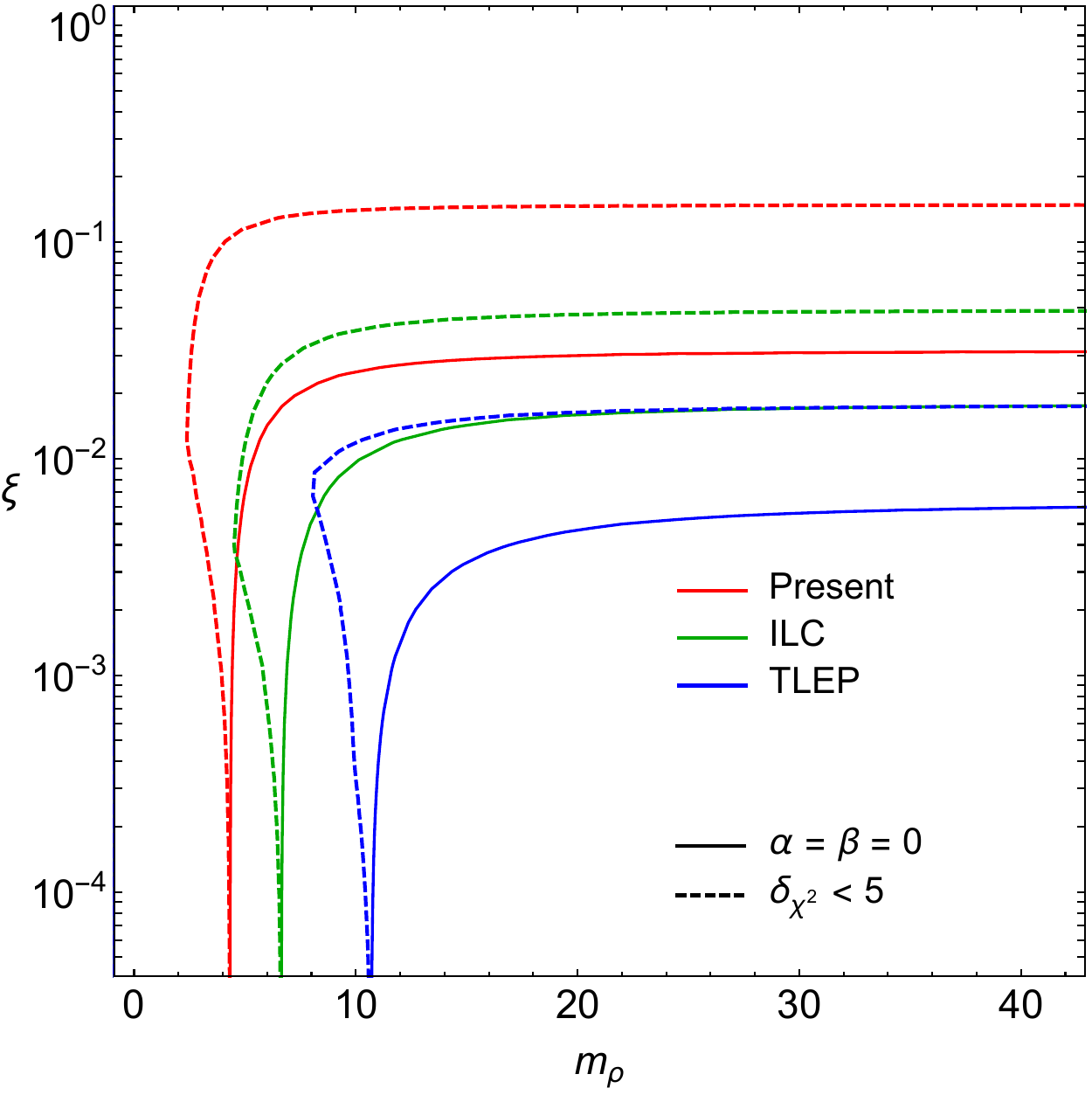}
\caption{\small Constraints from EWPT in the $(m_{\rho},\xi)$ plane in the regions relevant for the LHC at $14\,$TeV (left) and FCC at $100\,$TeV (right). The different dashings correspond to different hypotheses on $\alpha$ and $\beta$ in eq.~\eqref{oblique}: solid for $\alpha=\beta=0$ and dashed for $\delta_{\chi^{2}}<5$. The red curve represent the current constraints while green and blue are projections for the ILC and TLEP respectively.}\vspace{-0.5cm}\label{fig:EWPT}
\end{center}
\end{figure*}

\l{sec:ewpt}
\label{EWPT}
\vspace{-2mm}
As mentioned in the Introduction, EWPT, and in particular the oblique parameters $\hat{S}$ and $\hat{T}$, set some of the strongest constraints on CH models. However, as we stressed before, they suffer from an unavoidable model dependence, so that incalculable UV contributions can substantially relax these constraints \cite{Grojean:2013qca}. We believe that presenting the corresponding exclusion contours in the previous plots without taking into account any possible UV contribution would lead to a wrong and too pessimistic conclusion. Therefore we parametrize the new physics contributions to $\hat{S}$ and $\hat{T}$ as
\be\l{oblique}
\bry{l}
\dst \Delta\hat{S}=\f{g^{2}}{96\pi^{2}}\xi\log\(\f{\Lambda}{m_{h}}\)+\f{m_{W}^{2}}{m_{\rho}^{2}}+\alpha\f{g^{2}}{16\pi^{2}}\xi\,,\\
\dst \Delta\hat{T}=-\f{3g^{\prime\,2}}{32\pi^{2}}\xi\log\(\f{\Lambda}{m_{h}}\)+\beta\f{3y_{t}^{2}}{16\pi^{2}}\xi\,,
\ery
\ee
where the first terms represent the IR contributions due to the Higgs coupling modifications \cite{Barbieri:2007bh}, the second term in $\Delta\hat{S}$ comes from tree-level exchange of vector resonances and the last terms parametrize short distance effects. The scale $\Lambda$ in eq.~\eqref{EWPT} represents the scale of new physics, which we set to $\Lambda=4\pi f$. We could instead use $m_{\rho}$ to parametrize this scale, however, here we have the situation in mind where $m_{\rho}$ could be lighter than the typical resonances scale, or the cut-off scale, and our choice maximises the NP effect, leading to a more conservative bound. Moreover, being the sensitivity to this scale logarithmic, the final result only has a mild sensitivity on this choice. The coefficients $\alpha$ and $\beta$ are of order one and could have either sign \cite{Grojean:2013qca}. In the literature, a constant positive contribution to $\Delta\hat{T}$ has often been assumed to relax the constraints from EWPT \cite{Contino:2013un,Pappadopulo:2013wt}. However, 
the finite UV contributions of the form of the last terms in eq.~\eqref{oblique} arising from loops of heavy fermionic resonances always depend on $\xi$, significantly changing the EW fit compared to a constant contribution. In order to show realistic constraints from EWPT, we define a $\chi^{2}$ as a function of $\xi,m_{\rho},\alpha,\beta$, i.e.~$\chi^{2}(\xi,m_{\rho},\alpha,\beta)$, and compute $95\%\,$CL exclusion contours in the $(m_{\rho},\xi)$ plane marginalising over $\alpha$ and $\beta$. In order to control the level of cancellation in the $\chi^{2}$ due to the contribution of the UV terms, we define the parameter
\be
\delta_{\chi^{2}}=\f{\chi^{2}(\xi,m_{\rho},\alpha=0,\beta=0)}{\chi^{2}(\xi,m_{\rho},\alpha,\beta)}\,.
\ee 
In figure \ref{fig:EWPT} we show contours for $\alpha=\beta=0$ and $\delta_{\chi^{2}}<5$, which corresponds to a mild $20\%$ cancellation.
The marginalisation over $\alpha$ and $\beta$ is performed by scanning over them in a logarithmically symmetric interval $(1/3,3)$ for each point in the $(m_{\rho},\xi)$ plane. The dependence on the chosen interval is very mild once the constraints on $\delta_{\chi^{2}}$ are imposed.\footnote{We checked that the cancellation defined through the parameter $\delta_{\chi^{2}}$ gives comparable results as the cancellation defined in terms of the number of points satisfying the $95\%\,$CL bound over the number of points that do not satisfy it (or, in other words, the number of points falling within the $95\%$ CL ellipse in the $(\hat{S}$,$\hat{T})$ plane over the number of points falling outside).} For comparison with the future reach on Higgs couplings and direct resonance production, we show the constraints from EWPT with currently available data \cite{Baak:2012kk} together with the expectation at the ILC \cite{Baak:2014ora} and TLEP \cite{Baak_slides}. While currently masses below $\sim4.5\,$TeV are excluded for weak coupling (small $\xi$ in the plot) at $\alpha=\beta=0$, this bound will move to $\sim6.5\,$TeV and $\sim10.5\,$TeV at the ILC and TLEP respectively. For large values of $g_\rho$ (large $\xi$ in the plot), the bounds become more stringent. For now, at $\alpha=\beta=0$, EWPT exclude $\xi$-values above a few percent independently of $m_{\rho}$. While ILC only brings an improvement of a factor of two or three, TLEP is expected to reach a few per mille in $\xi$. All these bounds relax significantly when adding a non-vanishing UV contribution $\alpha,\beta\neq 0$ even for small values of $\delta_{\chi^{2}}$, i.e.~for not so un-natural cancellations in $\Delta\hat{S}$ and $\Delta\hat{T}$ induced by the UV contributions. In particular, vector masses down to $\sim 2-3\,$TeV can still be allowed for $\xi$ in the percent region, corresponding to $g_{\rho}$ couplings of order one. Moreover, the aforementioned few percent limit on $\xi$, independently of $m_{\rho}$, gets relaxed roughly to $10-15\%$. Finally notice that the expected relaxed constraint at ILC excludes larger $m_{\rho}$ values (up to $\sim 6.5\,$TeV) than LEP at $\alpha=\beta=0$ for small values of $g_{\rho}$, while giving a comparable constraint on $\xi$. Only TLEP will be able to push the relaxed bound below the actual LEP strict bound, still improving the limit on $\xi$ by only a factor of two. This gives an idea of the strong impact that UV contributions can give to the EWPT constraints and of their model dependence. We believe that the relaxed bounds that we show in figure \ref{fig:EWPT} represent a more realistic picture of the status of EWPT in CH models.

\section{Conclusions}\l{sec:conclusion}

We studied the complementarity of direct and indirect searches for the exploration of the CH scenario at the LHC and future colliders, by taking vector triplet production as a representative direct signature and Higgs coupling modifications as representative indirect constraints. The result, reported in section \ref{sec:results}, is that the relative discriminating power of the two search strategies crucially depends on the strength of the resonance coupling $g_\rho$: a weak coupling favours direct searches while strong coupling prefers indirect measurements. The threshold values of $g_\rho$ which set the boundary between the two regions are quantified in a comparison between different leptonic and hadronic collider options. The results indicate complementarity and do not allow us to draw a sharp conclusion on which strategy would be more effective because we do not have clear indications on the expected coupling strength. Even when dealing with a strongly--interacting microscopic theory the effective resonance coupling may well be weak for a large number of colours of the underlying strong interactions. Furthermore weakly coupled CH models are easily constructed as extra--dimensional holographic theories. Based on phenomenological considerations, two contradictory arguments could be made in favour of a strong or weak effective coupling. If we assume the level of fine--tuning in the theory to be exclusively controlled by $\xi=v^2/f^2$, i.e.~by how much the Higgs VEV is reduced with respect to the generic expectation $v\sim f$ by adjusting the parameters in the Higgs potential, we would prefer $f$ as small as possible and $g_\rho$ large to make the resonance scale $g_\rho f$ avoid EWPT constraints. This was the pattern we originally had in mind for CH theories. However it was subsequently realised, also because the Higgs boson turned out to be light, that the tuning also depends on the resonance scale $m_\rho =g_\rho f$, pushing us back to the small $g_\rho$ region. Actually, the tuning is not directly controlled by the mass of the vector resonance $m_\rho$, but instead by the one of the top partners $m_\Psi$. However there is no reason to expect a large gap between the two scales and only a mild accidental numerical separation seems tolerable. Given a value of $m_\Psi/f=g_\Psi\sim 2 $ for a light enough Higgs with moderate fine--tuning, it would be surprising to have $g_\rho$ much above $4$ or $5$. Composite Higgs models implementing the Twin Higgs protection \cite{Chacko:2005pe} for the Higgs potential might further change our expectations since in this case the tuning is disentangled from the resonance scale and the large $g_\rho$ regime is favoured again. Indirect searches are thus the most effective in the Twin Composite Higgs scenarios, at least in comparison with the direct heavy vector signatures we considered here. Better direct tests of the Twin CH most likely exist and need to be studied for a robust assessment.

At the technical level, we estimated the reach of direct searches by extrapolating the current $8\,$TeV limits based on luminosity rescaling as described in section \ref{sec:extrapolation}. This is meant to be a first estimate of the reach of future colliders, to be validated with detailed simulations. In the case of the FCC, the lack of detailed information on the detectors which might be employed clearly prevents a more detailed assessment for the time being. Conversely, the study of signals like the one we discussed here will itself contribute to the design of the detector. As far as indirect searches are concerned, we considered Higgs coupling modifications and, in section \ref{sec:ewpt}, the impact of current and future EWPT. Other indirect signatures should be added, among which precision measurements at lepton colliders other than the oblique $S$ and $T$ corrections and possible precision studies at hadron colliders. Clearly, hadron colliders are intrinsically less precise, but they produce hard reactions where the effects of Higgs compositeness might be enhanced. These consideration might apply, for example, to the $WW$ scattering process.

\section*{Acknowledgements}

We would like to thank Michelangelo Mangano for encouragement and useful conversations. R.T. and A.W.~thank Stefania De Curtis for discussion. A.T.~acknowledges support from the Cluster of Excellence {\it Precision Physics, Fundamental Interactions and Structure of Matter} (PRISMA -- EXC 1098). The work of R.T. was supported by the ERC Advanced Grant no.~267985 {\it DaMeSyFla} and by the Italian Ministero dell'Universit\'a e della Ricerca Scientifica under the PRIN Bando 2010-2011 no.~2010YJ2NYW$\_$003. A.W.~acknowledges the MIUR-FIRB grant \sloppy\mbox{RBFR12H1MW}. Part of the work has been completed in the context of the FCC programme at CERN and the {\it What Next} initiative by the Italian INFN.

\appendix

\section{A simple check of the extrapolation procedure}
\label{appendix}

\begin{figure*}[t!]
\begin{center}
\includegraphics[scale=0.35]{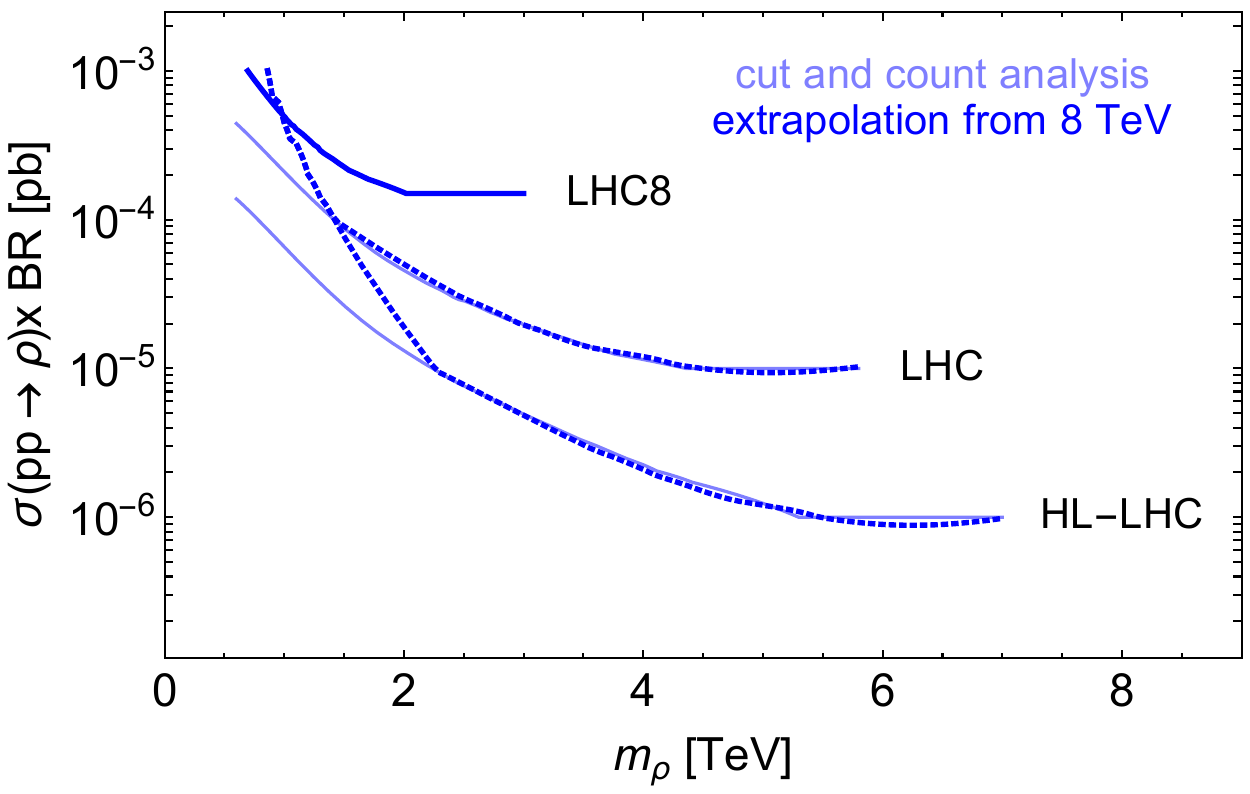}
\hspace{0.5cm}\includegraphics[scale=0.35]{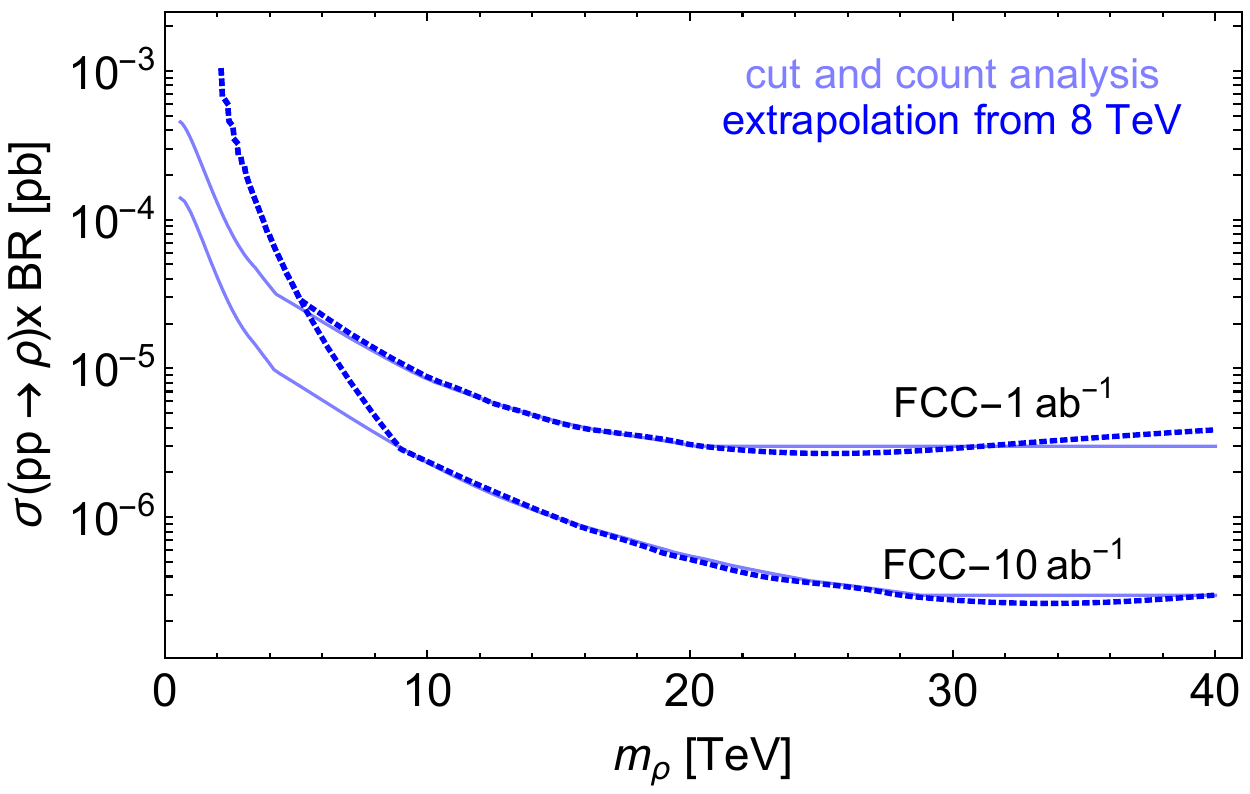}\vspace{-0.2cm}
\caption{\small Bounds on $\sigma\times \text{BR}$ from LHC at $8\,$TeV with $20\,$fb$^{-1}$ (LHC8)(thick, solid) and extrapolations to LHC at $14\,$TeV with $300\,$fb$^{-1}$ (LHC) and $3\,$ab$^{-1}$ (HL-LHC) and to FCC at $100\,$TeV with $1\,$ab$^{-1}$ and $10\,$ab$^{-1}$ (dark blue, dotted). Light blue lines represent the corresponding bounds obtained from a cut-and-count analysis.}\label{fig:extrapolationcheck}\vspace{-0.4cm}
\end{center}
\end{figure*}

We validated our extrapolation procedure described in section \ref{sec:extrapolation} against a simple cut-and-count analysis for di-lepton searches. The cut-and-count analysis is based on a di-lepton background simulation performed with {\sc{MadGraph5}} \cite{Alwall:2014hca} in the relevant invariant mass regions for an $8$, $14$ and $100\,$TeV collider. Counting events within an invariant mass window of $m_\rho \pm 0.1 m_\rho$ allows us to extract an exclusion limit on $\sigma \times$BR for each collider and luminosity configuration based solely on the background estimate. 
In parallel, we extrapolated the $8\,$TeV bound so obtained to higher energies and luminosities with the procedure outlined in section \ref{sec:extrapolation}. Exclusion limits from both methods are shown in figure \ref{fig:extrapolationcheck}. The thick solid blue curve depicts the $8\,$TeV bound obtained from the cut-and-count analysis which has been used for extrapolation, shown by the dotted blue lines. Thin lines in light blue represent cut-and-count limits for larger energies and luminosities. As can be seen, there is a perfect agreement at high masses. Of course, this is due to the fact that we use the same cut-and-count analysis for each collider configuration. More statistically refined analyses from the experimental collaborations could affect our conclusions. Here, however, it serves as a proof of principle. 
Since the background dominates in the high mass region, the limit scales linearly with the integrated luminosity. The scaling changes smoothly to the square root of the luminosity in the intermediate mass range. The extrapolation procedure fails for very low masses. As discussed at the end of section \ref{sec:extrapolation}, this is due to the fact that the $8\,$TeV bound starts at a certain lowest mass. The extrapolated low mass region is obtained from this lowest mass point and particularly small integrated luminosities which is not a reliable bound, as can be seen.

\bibliographystyle{mine}
\bibliography{draft}

\end{document}